\documentclass[a4paper,11pt]{article}
\usepackage{pos}
\usepackage{subfigure}
\usepackage{lineno}
\usepackage{xspace}
\usepackage{hyperref}
\usepackage{graphicx}
\usepackage{lineno}
\usepackage{color}

\title{Ground and excited quarkonium states as probes of MPI in small systems with ALICE}
\ShortTitle{Ground and excited quarkonium states as probes of MPI in small systems with ALICE}

\author*[a]{Theraa TORK}

\affiliation[a]{Laboratoire de Physique des 2 Infinis, Irène Joliot-Curie,\\
 Orsay, France}

\emailAdd{theraa.tork@cern.ch}

\abstract
{Quarkonia represent excellent tools for understanding the role of multiparton interactions (MPI) in small systems, i.e., pp and p-A collisions. Probing MPI with quarkonia can be done directly by looking at quarkonium associated production, or indirectly by studying the multiplicity dependence of quarkonium production. In these proceedings, the results from the ALICE experiment on direct and indirect MPI probes are discussed. The self-normalized yields of charmonium production as a function of the charged-particle multiplicity in pp and p--Pb collisions at \mbox{$\s$ = 13 TeV} and \mbox{$\snn$ = 8.16 TeV}, respectively, are discussed. Additionally, the corresponding measurements of the charmonium self-normalized excited-to-ground state ratio as a function of the charged-particle multiplicity are reported. The new measurement of the double $\Jpsi$ production at forward rapidity in pp collisions at \mbox{$\s$ = 13} TeV is presented. Finally, the multiplicity dependence of the bottomonium production in pp collisions at \mbox{$\s$ = 13 TeV} is discussed. The comparison to the available theoretical calculations is reported.}
\FullConference{%
  41st International Conference on High Energy physics - ICHEP2022\\
  6-13 July, 2022\\
  Bologna, Italy
}

\begin{document}
%

\newcommand{\pp}           {pp\xspace}
\newcommand{\ppbar}        {\mbox{$\mathrm {p\overline{p}}$}\xspace}
\newcommand{\XeXe}         {\mbox{Xe--Xe}\xspace}
\newcommand{\PbPb}         {\mbox{Pb--Pb}\xspace}
\newcommand{\pA}           {\mbox{pA}\xspace}
\newcommand{\pPb}          {\mbox{p--Pb}\xspace}
\newcommand{\Pbp}          {\mbox{Pb--p}\xspace}
\newcommand{\AuAu}         {\mbox{Au--Au}\xspace}
\newcommand{\dAu}          {\mbox{d--Au}\xspace}

\newcommand{\s}            {\ensuremath{\sqrt{s}}}
\newcommand{\snn}          {\ensuremath{\sqrt{s_{\mathrm{NN}}}}}
\newcommand{\pt}           {\ensuremath{p_{\rm T}}\xspace}
\newcommand{\meanpt}       {$\langle p_{\mathrm{T}}\rangle$\xspace}
\newcommand{\ycms}         {\ensuremath{y_{\rm cms}}\xspace}
\newcommand{\ylab}         {\ensuremath{y_{\rm lab}}\xspace}
\newcommand{\etacms}       {\ensuremath{\eta_{\rm cms}}\xspace}
\newcommand{\etalab}       {\ensuremath{\eta_{\rm lab}}\xspace}
\newcommand{\etarange}[1]  {\mbox{$\left | \eta \right |~<~#1$}}
\newcommand{\yrange}[1]    {\mbox{$\left | y \right |~<~#1$}}
\newcommand{\dndy}         {\ensuremath{\mathrm{d}N_\mathrm{ch}/\mathrm{d}y}\xspace}
\newcommand{\dndeta}       {\ensuremath{\mathrm{d}N_\mathrm{ch}/\mathrm{d}\eta}\xspace}
\newcommand{\avdndeta}     {\ensuremath{\langle\dndeta\rangle}\xspace}
\newcommand{\dNdy}         {\ensuremath{\mathrm{d}N_\mathrm{ch}/\mathrm{d}y}\xspace}
\newcommand{\Npart}        {\ensuremath{N_\mathrm{part}}\xspace}
\newcommand{\Ncoll}        {\ensuremath{N_\mathrm{coll}}\xspace}
\newcommand{\dEdx}         {\ensuremath{\textrm{d}E/\textrm{d}x}\xspace}
\newcommand{\RpPb}         {\ensuremath{R_{\rm pPb}}\xspace}

\newcommand{\nineH}        {$\sqrt{s}~=~0.9$~Te\kern-.1emV\xspace}
\newcommand{\thirteen}        {$\sqrt{s}~=~13$~Te\kern-.1emV\xspace}
\newcommand{\seven}        {$\sqrt{s}~=~7$~Te\kern-.1emV\xspace}
\newcommand{\twoH}         {$\sqrt{s}~=~0.2$~Te\kern-.1emV\xspace}
\newcommand{\twosevensix}  {$\sqrt{s}~=~2.76$~Te\kern-.1emV\xspace}
\newcommand{\five}         {$\sqrt{s}~=~5.02$~Te\kern-.1emV\xspace}
\newcommand{\twosevensixnn}{$\sqrt{s_{\mathrm{NN}}}~=~2.76$~Te\kern-.1emV\xspace}
\newcommand{\fivenn}       {$\sqrt{s_{\mathrm{NN}}}~=~5.02$~Te\kern-.1emV\xspace}
\newcommand{\LT}           {L{\'e}vy-Tsallis\xspace}
\newcommand{\GeVc}         {Ge\kern-.1emV/$c$\xspace}
\newcommand{\MeVc}         {Me\kern-.1emV/$c$\xspace}
\newcommand{\TeV}          {Te\kern-.1emV\xspace}
\newcommand{\GeV}          {Ge\kern-.1emV\xspace}
\newcommand{\MeV}          {Me\kern-.1emV\xspace}
\newcommand{\GeVmass}      {Ge\kern-.2emV/$c^2$\xspace}
\newcommand{\MeVmass}      {Me\kern-.2emV/$c^2$\xspace}
\newcommand{\lumi}         {\ensuremath{\mathcal{L}}\xspace}

\newcommand{\ITS}          {\rm{ITS}\xspace}
\newcommand{\TOF}          {\rm{TOF}\xspace}
\newcommand{\ZDC}          {\rm{ZDC}\xspace}
\newcommand{\ZDCs}         {\rm{ZDCs}\xspace}
\newcommand{\ZNA}          {\rm{ZNA}\xspace}
\newcommand{\ZNC}          {\rm{ZNC}\xspace}
\newcommand{\SPD}          {\rm{SPD}\xspace}
\newcommand{\SDD}          {\rm{SDD}\xspace}
\newcommand{\SSD}          {\rm{SSD}\xspace}
\newcommand{\TPC}          {\rm{TPC}\xspace}
\newcommand{\TRD}          {\rm{TRD}\xspace}
\newcommand{\VZERO}        {\rm{V0}\xspace}
\newcommand{\TZERO}        {\rm{T0}\xspace}
\newcommand{\VZEROA}       {\rm{V0A}\xspace}
\newcommand{\VZEROC}       {\rm{V0C}\xspace}
\newcommand{\Vdecay} 	   {\ensuremath{V^{0}}\xspace}

\newcommand{\ee}           {\ensuremath{e^{+}e^{-}}} 
\newcommand{\pip}          {\ensuremath{\pi^{+}}\xspace}
\newcommand{\pim}          {\ensuremath{\pi^{-}}\xspace}
\newcommand{\kap}          {\ensuremath{\rm{K}^{+}}\xspace}
\newcommand{\kam}          {\ensuremath{\rm{K}^{-}}\xspace}
\newcommand{\pbar}         {\ensuremath{\rm\overline{p}}\xspace}
\newcommand{\kzero}        {\ensuremath{{\rm K}^{0}_{\rm{S}}}\xspace}
\newcommand{\lmb}          {\ensuremath{\Lambda}\xspace}
\newcommand{\almb}         {\ensuremath{\overline{\Lambda}}\xspace}
\newcommand{\Om}           {\ensuremath{\Omega^-}\xspace}
\newcommand{\Mo}           {\ensuremath{\overline{\Omega}^+}\xspace}
\newcommand{\X}            {\ensuremath{\Xi^-}\xspace}
\newcommand{\Ix}           {\ensuremath{\overline{\Xi}^+}\xspace}
\newcommand{\Xis}          {\ensuremath{\Xi^{\pm}}\xspace}
\newcommand{\Oms}          {\ensuremath{\Omega^{\pm}}\xspace}
\newcommand{\degree}       {\ensuremath{^{\rm o}}\xspace}

\newcommand{\Jpsi}{\rm{J}/\psi}
\newcommand{\psit}{\psi(2{\rm S})}
\newcommand{\psip}{\psi(2{\rm S})}
\newcommand{\psipOJpsi}{\ensuremath{\psip\mbox{-over-}\Jpsi}}
\newcommand{\abs}[1]{\left\vert #1 \right\vert}
\newcommand{\ave}[1]{\langle #1 \rangle}
\newcommand{\zv}{z_{\rm vtx}}
\newcommand{\Ntr}{N_{\rm tracklet}}
\newcommand{\avNtr}{\langle N_{\rm tracklet} \rangle}
\newcommand{\Ntrc}{N_{\rm tracklet}^{\rm corr}}
\newcommand{\Ncorr}        {\ensuremath{N^\mathrm{corr}}\xspace}
\newcommand{\Ae}{A\varepsilon}
\newcommand{\mumu}{\mu^+\mu^-}
\newcommand{\BR}{\rm{BR}_{\Jpsi\rightarrow\mumu}}

\newcommand{\Ups}[1]{\Upsilon(#1{\rm nS})}
\newcommand{\UpsGr}[1]{\Upsilon(#1{\rm 1S})}
\newcommand{\UpsTwo}[1]{\Upsilon(#1{\rm 2S})}
\newcommand{\UpsThre}[1]{\Upsilon(#1{\rm 2S})}

\newcommand{\eightnn}       {$\sqrt{s_{\mathrm{NN}}}~=~8.16$~Te\kern-.1emV\xspace}
\newcommand{\AAcoll}          {\mbox{A--A}\xspace}

\maketitle


\section{Introduction}
\noindent In ultrarelativistic heavy-ion collisions, a deconfined state of QCD matter, made of free quarks and gluons, called the quark-gluon plasma (QGP) is expected to be formed.
\noindent Several hard and soft processes are used to probe such a medium, e.g., quarkonia for hard probes. Quarkonia are bound states of heavy quark anti-quark pairs, e.g., charmonia (c$\Bar{\rm c}$) and bottomonia (b$\Bar{\rm b}$). Quarkonia are expected to experience the full evolution of the collision, due to the early production of their constituent quarks, and interact strongly with the medium. The study of quarkonium production in small systems like pp and pA collisions, where no QGP effects are expected, is essential as a reference for A-A collisions. However, several QGP-like behaviors are observed in small systems for high multiplicity events, such as the non-zero elliptic flow of identified hadrons measured via long-range angular correlations in p--Pb collisions at $\sqrt{s_{\rm NN}}$ = 5.02 TeV  \cite{ALICE:2013snk} and the enhanced production of multi-strange hadrons in pp collisions at $\sqrt{s}$ = 7 TeV \cite{ALICE:2016fzo}. These features are interpreted as signs of the QGP formation in A-A collisions. Therefore, it is important to characterize the initial state of hadronic collisions and understand the mechanisms that can contribute to the high charged-particle multiplicity events. The non-linearities originating from multiparton interactions (MPI) is one of the main scenarios proposed to describe these findings. MPI are defined as several parton-parton interactions that take place in a single nucleon-nucleon collision. One can probe MPI by studying the quarkonium associated production or the quarkonium production as a function of the charged-particle multiplicity. The first analysis represents a direct probe to MPI. However, it requires a large luminosity to be performed. The second analysis is an indirect probe to MPI, as the multiplicity of charged particles is correlated with the number of MPI in the event. The results of the multiplicity dependence of quarkonium production from the ALICE experiment in pp and p--Pb collisions at \mbox{$\s$ = 13 TeV} and $\snn$ = 8.16 TeV, respectively, are discussed. In addition, the charmonium associated production results in pp collisions at \mbox{$\s$ = 13 TeV} is reported.         
     

\section{Experimental setup}
 \noindent A Large Ion Collider Experiment (ALICE) is a general purpose detector devoted to the study of A-A collisions, as well as pA and pp collisions at the CERN LHC. A detailed description of the ALICE experiment and its performance can be found in \cite{ALICE:2014sbx}. The detector consists of 18 different sub-detectors. In these proceedings the central barrel detectors at midrapidity, |$y\rm{_{lab}}$| < 1, are used to reconstruct the quarkonium states in their decay channel into dielectrons and the charged-particle multiplicity. The Muon spectrometer at forward rapidity, 2.5 < $y\rm{_{lab}}$ < 4.0, is optimised to reconstruct the quarkonium states in their decay channel into dimuons. The ALICE results presented here are for the inclusive quarkonium production, i.e, no separation is made between prompt quarkonia and those from b-hadron decays, the so-called non-prompt component.   

\section{Results}
\subsection{J/$\psi$ pair production in pp collisions at $\sqrt{s}$ = 13 TeV}
\begin{figure}[!htbp]
 \centering
    \subfigure[J/$\psi$ pair production]{
    \includegraphics[width =0.48\textwidth]{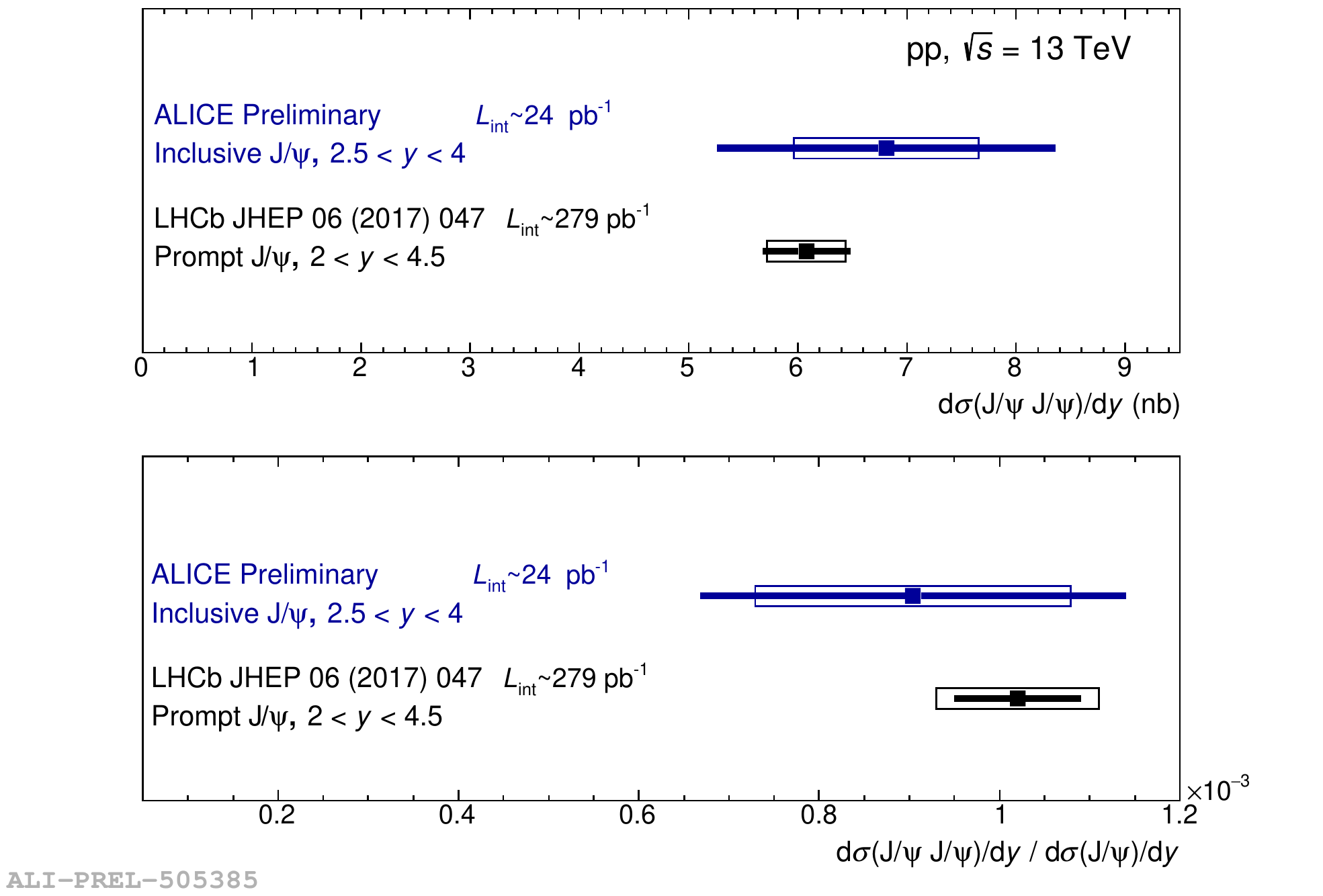}

    \label{fig:doubleJPsi}
  }
    \qquad 
    \subfigure[J/$\psi$ self-normalized yields at midrapidity]{
    \includegraphics[width =0.43\textwidth]{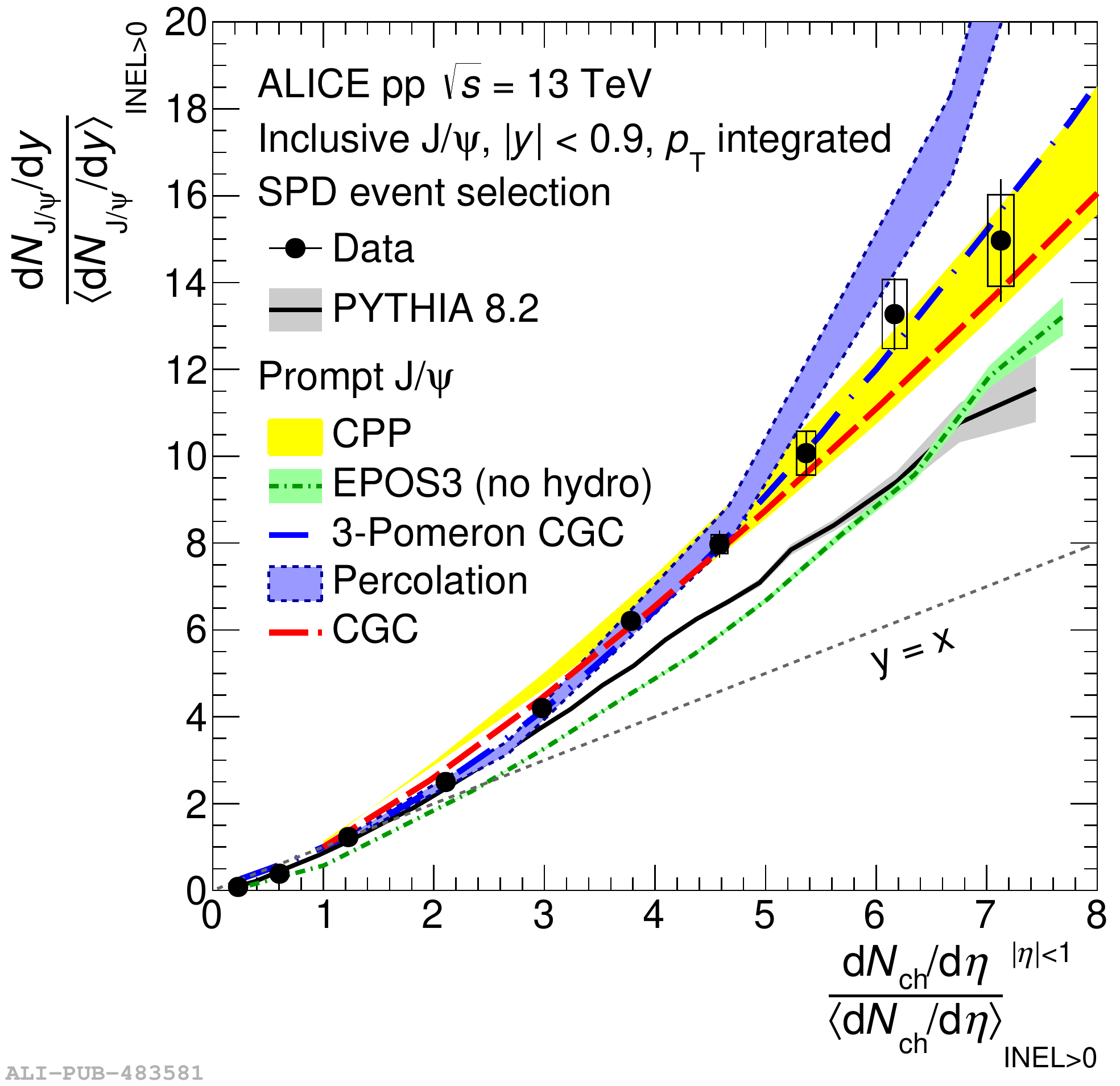}

    \label{fig:JPSi_mid}
  }
    \caption{(a) Inclusive J/$\psi$ pair production cross section (top panel), and ratio of double-to-single inclusive J/$\psi$ production cross sections (bottom panel), in pp collisions at $\sqrt{s}$ = 13 TeV. Results are compared with LHCb measurements for prompt J/$\psi$ production \cite{LHCb:2016wuo}. (b) J/$\psi$ self-normalized yields as a function of the self-normalized charged-particle multiplicity \cite{ALICE:2020msa}. Both quantities are measured at midrapidity. The results are compared with several theoretical models \cite{Kopeliovich:2013yfa,Levin:2019fvb,Ma:2014mri,Ferreiro:2012fb,Sjostrand:2014zea,Drescher:2000ha}.
    }
    \label{fig:pp_odubleandmultiplicity}
\end{figure}

\noindent The J/$\psi$ pair production is measured in ALICE exploiting the data from pp collisions at \mbox {$\s$ = 13 TeV}. The measurement represents a direct probe to MPI, as it is sensitive to the single and double-parton hard scatterings. Moreover, such a measurement allows for a better understanding of the single J/$\psi$ production mechanism \cite{Szczurek:2012kt}. 
\noindent The result in Fig.\ref{fig:doubleJPsi} shows the inclusive J/$\psi$ pair production cross section in the top panel and the ratio of the double-to-single J/$\psi$ production cross section in the bottom panel. Each measurement is compared to the corresponding prompt result measured by the LHCb collaboration in a slightly larger rapidity range \cite{LHCb:2016wuo}. The ALICE and LHCb results are consistent within uncertainties.  

\subsection{The multiplicity dependence of self-normalized charmonium production}
\noindent Fig.\ref{fig:JPSi_mid} shows the self-normalized J/$\psi$ yields as a function of the charged-particle multiplicity, both measured at midrapidity, in pp collisions at $\s$ = 13 TeV \cite{ALICE:2020msa}. The measured yields exhibit a faster than linear increase with increasing multiplicity. The measurement is compared to several theoretical models which attribute the observed behavior to different initial and/or final state effects. The theoretical calculations from the coherent particle production approach (CPP) \cite{Kopeliovich:2013yfa}, the  color glass condensate model (CGC) \cite{Ma:2014mri}, and 3-pomeron CGC \cite{Levin:2019fvb} describe the results quantitatively in the probed multiplicities. The percolation model \cite{Ferreiro:2012fb} describes the observed trend of the self-normalized J/$\psi$ yields up to 5 times the average multiplicity. PYTHIA 8.2 \cite{Sjostrand:2014zea} and EPOS3 \cite{Drescher:2000ha} event generators predict a faster than linear increase, however the overall slope of the measurement is underestimated.


\noindent The comparison between forward and midrapidity self-normalized J/$\psi$ yields as a function of the charged-particle multiplicity in pp collisions at $\s$ =13 TeV is reported in \cite{ALICE:2021zkd}. The trend of the forward J/$\psi$ measurement is compatible with a linear increase within uncertainties. This increase is less rapid than the one in the result at midrapidity. The measurement is compatible qualitatively with the theoretical computations from 3-pomeron CGC, percolation, and CPP models within uncertainties, while EPOS3 and PYTHIA 8.2 event generators underestimate the trend at high multiplicity. The result for the $\psip$ self-normalised yields, measured at forward rapidity, is done exploiting the pp dataset at $\s$ = 13 TeV as reported in \cite{ALICE:2022gpu}. In Fig.\ref{fig:psi2s}, an increase in the self-normalized $\psip$ yields as a function of the charged-particle multiplicity is observed. 
 This increase is compatible with a linear trend within uncertainties, as also predicted by \mbox{PYTHIA 8.2} calculations with and without color reconnection. \mbox{PYTHIA 8.2} predictions start to deviate from linearity above 5 times the average charged-particle multiplicity. The ratio of the self-normalized $\psip$-to-J/$\psi$ yields as a function of the charged-particle multiplicity is also presented in \cite{ALICE:2022gpu}. Such a measurement is important to disentangle any final-state effects at play, as most of the initial-state effects cancel out in the ratio. The trend of the result is compatible with a flat distribution suggesting a similar behavior as a function of the charged-particle multiplicity for the $\psip$ and J/$\psi$ states within uncertainties. This trend is reproduced by \mbox{PYTHIA 8.2} calculations. On the other hand, the comover model calculations \cite{Ferreiro:2014bia} suggest a stronger suppression of the excited state compared to the ground state at high multiplicities. In the comover model, the quarkonia are expected to be produced during the hard process of the collision and then interact with the soft particles that are comoving with them, leading to a dissociation of quarkonia. The dissociation probability will depend on the density of soft particles and the size of quarkonia (the larger the size, the stronger the suppression). The measurement is compatible with the comover model within uncertainties; however, the available statistics do not allow disentangling any possible final state effects at play.
 \\
\noindent In order to understand the influence of the nuclear environment on the production of charmonia, ALICE measured the J/$\psi$ and $\psip$ self normalized-yields as a function of the charged-particle multiplicity in p--Pb collisions at $\snn$ = 8.16 TeV \cite{ALICE:2020eji,ALICE:2022gpu}. The J/$\psi$ yields exhibit an increase faster (slower) than a linear at backward (forward) rapidity with the increasing charged-particle multiplicity. The results are compatible with EPOS3 event generator calculations within uncertainties. A similar behavior for $\psip$ self-normalized yields as a function of the charged-particle multiplicity is observed at forward and backward rapidity in p--Pb collisions at $\snn$ = 8.16 TeV, within the large experimental uncertainties. The percolation + comover + EPS09 model is compatible with the data within uncertainties \cite{Ferreiro:2014bia,Ferreiro:2012fb,Eskola:2009uj}. The self-normalized $\psip$-to-J/$\psi$ ratio as a function of the charged-particle multiplicity is compatible with unity within uncertainties. The comover model calculation suggests a stronger suppression of the $\psip$ with respect to the J/$\psi$, in both rapidity regions, due to the larger size of the $\psip$. A stronger suppression is expected at backward rapidity compared to the forward one, due to the larger density of soft particles produced. The comover model is compatible with data within the large experimental uncertainties.

 


\subsection{The multiplicity dependence of self-normalized bottomonium production }
\begin{figure}[!htbp]
 \centering
    \subfigure[$\psi(2S)$ self-normalised yield.]{
    \includegraphics[width =0.44\textwidth]{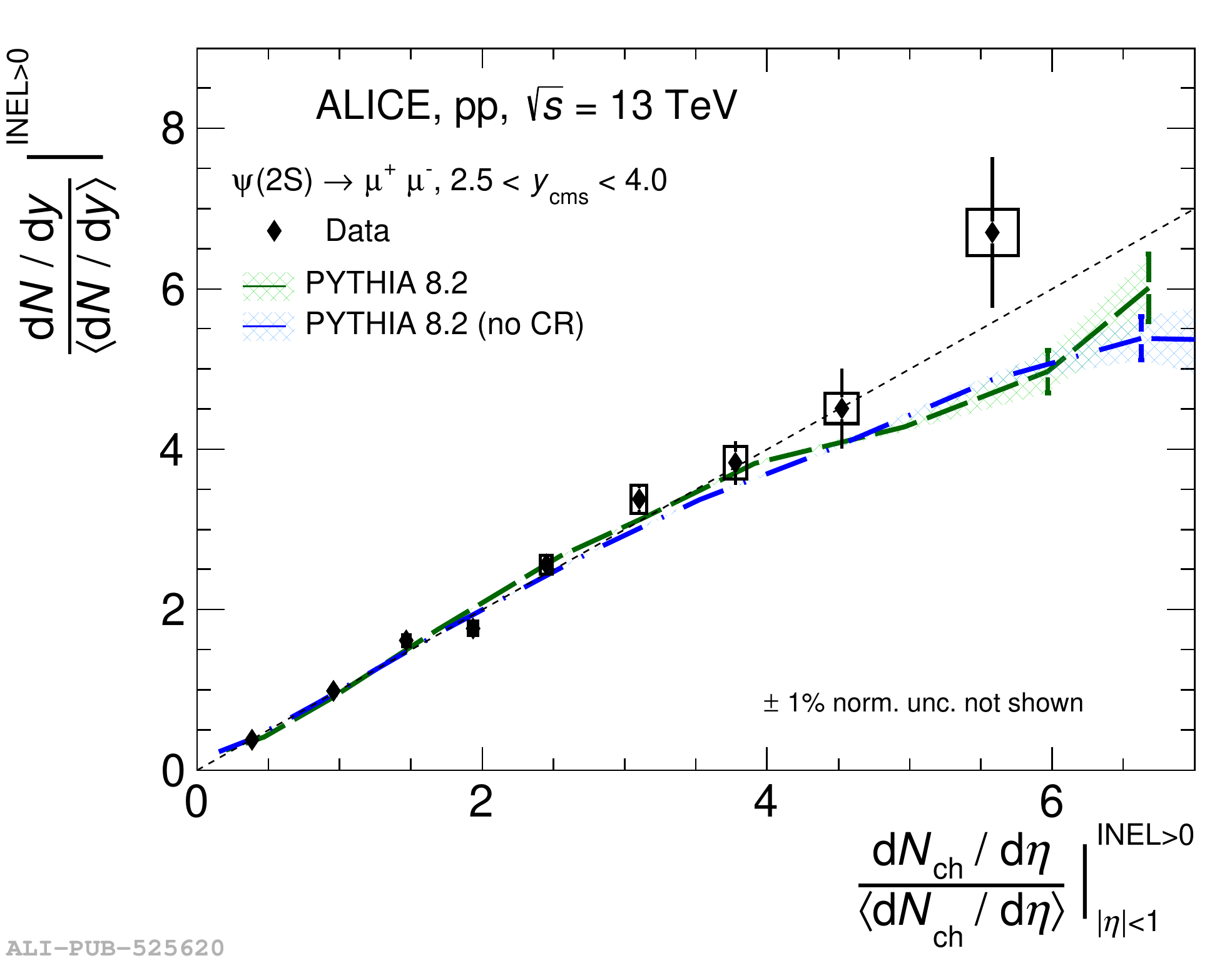}
    \label{fig:psi2s}
  }
    \qquad 
    \subfigure[$\UpsTwo$-to$\UpsGr$ self-normalized ratio]{
    \includegraphics[width =0.46\textwidth]{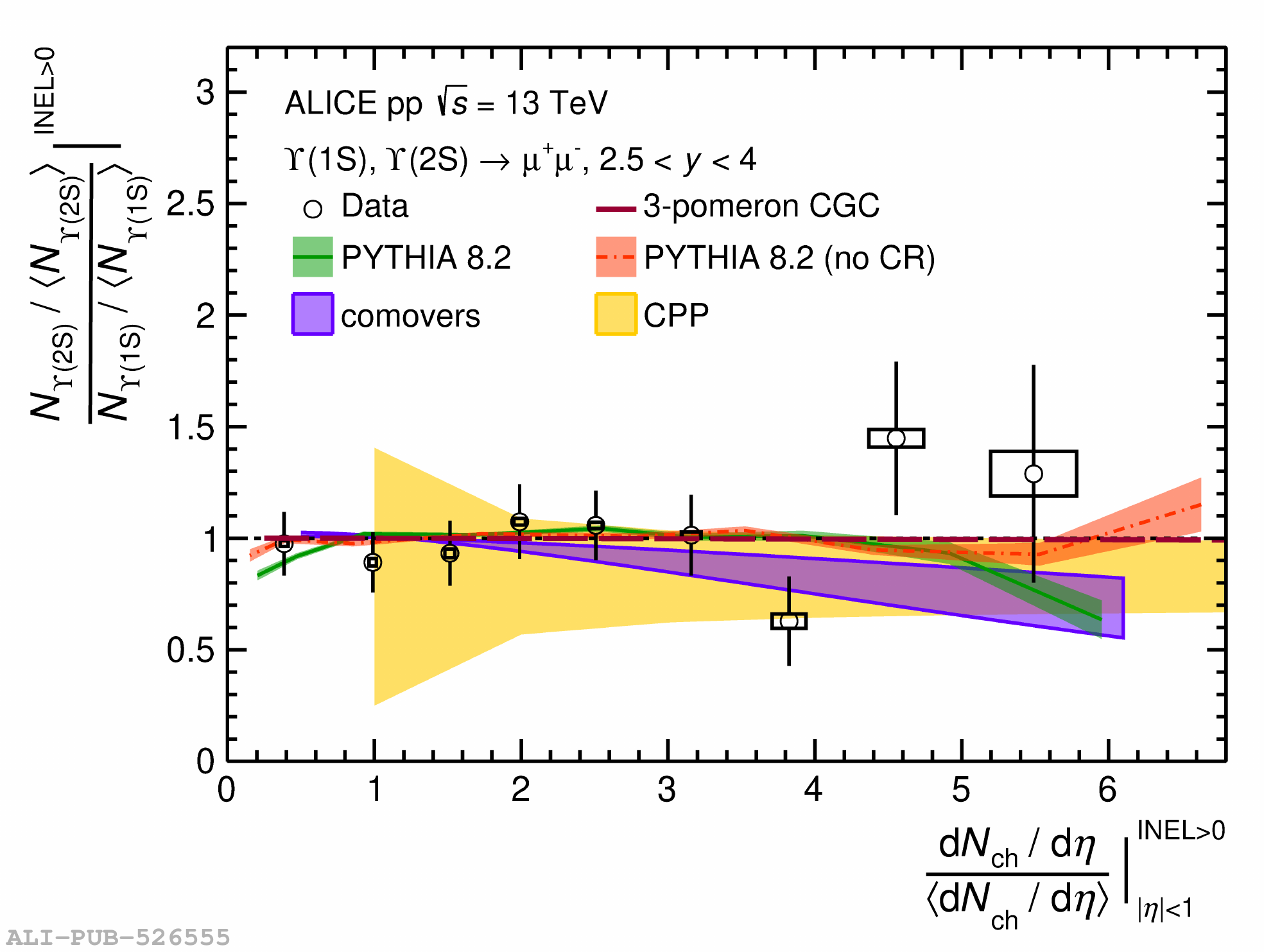}
    \label{fig:doubleY1SY2S}
}
     \caption{(a) Inclusive $\psi(2S)$ self-normalised yields, measured at forward rapidity, as a function of the self-normalised charged-particle multiplicity measured at midrapidity, in pp collisions at $\sqrt{s} = 13$~TeV\cite{ALICE:2022gpu}. Results are compared with predictions from \mbox{PYTHIA 8.2} \cite{Sjostrand:2014zea}. (b)$\UpsTwo$-to$\UpsGr$ self-normalized ratio as a function of the charged-particle multiplicity in pp collisions at $\s$ = 13 TeV \cite{ALICE:2022yzs}. The measurements are compared to theoretical model calculations from 3 pomeron-CGC approach \cite{Levin:2019fvb}, PYTHIA 8.2 event generator \cite{Sjostrand:2014zea}, comover model \cite{Ferreiro:2014bia}, coherent particle production (CPP) \cite{Kopeliovich:2013yfa}. 
    }
    \label{fig:bottomonia_doubleRatios}
\end{figure}

 \noindent Fig.\ref{fig:doubleY1SY2S} presents the $\UpsTwo$-to-$\UpsGr$ self-normalized ratios as a function of the charged-particle multiplicity \cite{ALICE:2022yzs}. The normalized ratio is compatible with unity within uncertainty up to 6 times the average multiplicity. The calculation from PYTHIA 8.2 event generator is compatible with the observed behaviour, independently of the considered color reconnection scenario, within uncertainties. The 3-pomeron CGC and CPP approaches predict a trend compatible with unity up to the probed multiplicity. The measurement is also compatible, within uncertainties, with the comover model calculation, which suggests a stronger suppression for the $\UpsTwo$ at high multiplicity compared to the ground state $\UpsGr$.   

\noindent The bottomonium-to-charmonium ground states self-normalised yields is also reported in \cite{ALICE:2022yzs}. The measurement implies no difference between the multiplicity dependence of the J/$\psi$ and $\UpsGr$ self-normalized yields. The predictions from the PYTHIA 8.2 event generator, 3-pomeron CGC, comover model and CPP approach are compatible with the result within uncertainties.  
 
  \section{Conclusion}
 \noindent ALICE measured the inclusive J/$\psi$ pair production cross section in pp collisions at \mbox{$\s$ =  13 TeV}. The result shows a good agreement within uncertainties with similar measurements performed by the LHCb experiment. The multiplicity dependence of J/$\psi$ self-normalised yields at midrapidity shows a faster than a linear increase in pp collisions at $\s$ = 13 TeV. The forward self-normalized J/$\psi$ and $\psip$ yields exhibit a linear increase as a function of the charged-particle multiplicity in pp and p--Pb collisions at $\s$ = 13 TeV and $\snn$ = 8.16 TeV, respectively. A similar trend is observed for the forward self-normalized $\Ups$ yields as a function of the charged-particle multiplicity for pp collisions at \mbox{$\s$ = 13 TeV}. The trend for both charmonium and bottomonium self-normalized yields is reproduced by several models, which include initial and/or final state effects, with MPI being included in the initia- state effects. The multiplicity dependence of the excited-to-ground quarkonium production shows a trend compatible with unity within uncertainties, independently of the quark flavor. The LHC Run 3 is expected to provide higher luminosity,thereby  allowing measurements as a function of the charged-particle multiplicity and to further constrain MPI modeling.

{
\footnotesize
\bibliographystyle{utphys}   
\bibliography{alice_pub,models,OtherCollaborations}
}
\end{document}